\documentclass[twocolumn]{aastex631}

\shorttitle{}
\shortauthors{Campbell et al.}
\graphicspath{{./}{}}

\begin{document}

\title{DKIST unveils the serpentine topology of quiet Sun magnetism in the photosphere}

\author[0000-0001-5699-2991]{Ryan J. Campbell}
\affiliation{Astrophysics Research Centre, School of Mathematics and Physics, Queen's University, Belfast, BT7 1NN, UK}

\author[0000-0001-8556-470X]{P. H. Keys}
\affiliation{Astrophysics Research Centre, School of Mathematics and Physics, Queen's University, Belfast, BT7 1NN, UK}

\author[0000-0002-7725-6296]{M. Mathioudakis}
\affiliation{Astrophysics Research Centre, School of Mathematics and Physics, Queen's University, Belfast, BT7 1NN, UK}

\author[0000-0001-6907-9739]{F. W\"{o}ger}
\affiliation{National Solar Observatory, 3665 Discovery Drive, Boulder, CO 80303, USA}

\author[0000-0002-7451-9804]{T. Schad}
\affiliation{National Solar Observatory, 22 ´Ōhi´a Kū Street, Pukalani, HI, 96768, USA}

\author[0000-0003-3147-8026]{A. Tritschler}
\affiliation{National Solar Observatory, 3665 Discovery Drive, Boulder, CO 80303, USA}

\author[0000-0002-5084-4661]{A. G. de Wijn}
\affiliation{High Altitude Observatory, National Center for Atmospheric Research, P.O. Box 3000, Boulder, CO 80307-3000, USA}

\author[0000-0003-3490-6532]{H. N. Smitha}
\affiliation{Max-Planck-Institut f\"ur Sonnensystemforschung, Justus-von-Liebig-Weg 3, 37077 G\"ottingen, Germany}

\author[0000-0001-7706-4158]{C. Beck}
\affiliation{National Solar Observatory, 22 ´Ōhi´a Kū Street, Pukalani, HI, 96768, USA}

\author[0000-0003-1746-3020]{D. J. Christian}
\affiliation{Department of Physics and Astronomy, California State University Northridge, 18111 Nordhoff Street, Northridge, CA 91330, USA}

\author[0000-0002-9155-8039]{D. B. Jess}
\affiliation{Astrophysics Research Centre, School of Mathematics and Physics, Queen's University, Belfast, BT7 1NN, UK}
\affiliation{Department of Physics and Astronomy, California State University Northridge, 18111 Nordhoff Street, Northridge, CA 91330, USA}

\author[0000-0003-3439-4127]{R. Erdélyi}
\affiliation{Solar Physics and Space Plasma Research Centre, School of Mathematics and Statistics, University of Sheffield, Sheffield S3 7RH, UK}
\affiliation{Department of Astronomy, E\"otv\"os Lor\'and University, P\'azm\'any P\'eter s\'et\'any 1/A, H-1117 Budapest, Hungary}


\begin{abstract}
{We present the first quiet Sun spectropolarimetric observations obtained with the Visible SpectroPolarimeter (ViSP) at the $4-$m Daniel K. Inouye Solar Telescope (DKIST). We recorded observations in a wavelength range that includes the magnetically sensitive Fe I $6301.5/6302.5$ $\mathrm{\AA}$ doublet. With an estimated spatial resolution of $0{\,}.{\!\!}{\arcsec}08$, this represents the highest spatial resolution full-vector spectropolarimetric observations ever obtained of the quiet Sun. We identified $53$ small-scale magnetic elements, including $47$ magnetic loops and $4$ unipolar magnetic patches, with linear and circular polarisation detected in all of them. Of particular interest is a magnetic element in which the polarity of the magnetic vector appears to change three times in only $400$ km and which has linear polarisation signals throughout. We find complex Stokes $V$ profiles at the polarity inversion lines of magnetic loops and discover degenerate solutions, as we are unable to conclusively determine whether these arise due to gradients in the atmospheric parameters or smearing of opposite polarity signals. We analyse a granule which notably has linear and circular polarisation signals throughout, providing an opportunity to explore its magnetic properties. On this small scale we see the magnetic field strength range from $25$ G at the granular boundary to $2$ kG in the intergranular lane (IGL), and sanity check the values with the weak and strong field approximations. A value of $2$ kG in the IGL is among the highest measurements ever recorded for the internetwork.}

\end{abstract}


\section{Introduction} \label{sec:intro}
 The discovery of transient, transverse 
 magnetism in the quiet Sun by \cite{lites1996} provided observational evidence suggesting that the ``magnetic carpet'' \citep{title1998} may not be predominantly longitudinal in nature. The short-lived linear polarisation, observed with the Zeeman effect, was found to be located at the edge of granules, between opposite polarity circular polarisation signals, but in the absence of any accompanying measurable Stokes $V$ signal at the polarity inversion line (PIL). Later, \cite{lites2008} used the spatial resolution afforded by the Hinode spacecraft and the magnetically sensitive $6302.5$~$\mathrm{\AA}$ line to unveil the spatially averaged transverse magnetic flux density ($55$~Mx~cm$^{-2}$) as five times greater than the longitudinal equivalent ($11$~Mx~cm$^{-2}$). However, circular polarisation remains much more prevalent than linear polarisation in this data. Long integrations with the Hinode/SpectroPolarimeter (SP) in sit-and-stare mode have revealed linear polarisation in as much as half of the field of view (FOV), but with compromised spatiotemporal resolution due to an integration time of $6.1$ minutes \citep{rubio2012}. Whether the vector magnetic field is predominantly horizontal (transverse) or vertical (longitudinal) in the solar photosphere is still debated, with the impact of photon noise on the retrieval of the magnetic inclination angle a source of controversy \citep{Borrero2011,danilovic2016}. Therefore, geometric approaches were also sought to circumvent this problem \citep[e.g.,][to name but a few]{stenflo2010,stenflo2013,shahin2014,lites2017}. The reader is directed to \cite{rubio2019} for a review.

The conversion and transport of quiet Sun magnetic energy in the photosphere to kinetic energy in the chromosphere and corona could have an important role to play in explaining the high temperature of the upper solar atmosphere \citep{Schrijver1998,bueno2004}. Statistical analyses of this phenomenon with the Zeeman effect suggest that magnetic bi-poles do not appear uniformly on the solar surface and there are regions of the very quiet Sun where no bi-poles seem to emerge \citep{marian2012}. \cite{marian2009} showed that $23\%$ of magnetic loops studied reached the chromosphere, perhaps driven by convective up-flows and magnetic buoyancy \citep{steiner2008}, on a timescale of $8$ minutes. \cite{gosic2021} demonstrated that small magnetic bi-poles that emerge in the photosphere with field strengths between $400$~G and $850$~G can reach the chromosphere, but on a much longer timescale of up to $1$ hour. \cite{wiegelmann2013} extrapolated photospheric magnetic field line measurements into the chromosphere and corona, and concluded that the energy released by magnetic reconnection could not fully account for the heating of these layers. However, the $110-130$~km resolution of their observations means this analysis does not account for small-scale braiding of magnetic field lines and their potential to drive heating via reconnection and dissipation, a mechanism which could explain the relative weakness of the small-scale magnetic field in the solar photosphere \citep{parker1972}.

Near infrared spectropolarimetric observations have a demonstrated ability to measure linear polarisation more effectively compared to photospheric diagnostics in the visible \citep{marian2008_mu,lagg2016,Campbell2021a}. Recently, \cite{campbell2023} revealed an internetwork region with a relatively large fraction of linear and circular polarisation ($23\%$ versus $60\%$) such that a majority of magnetised pixels displayed a clear transverse component of the magnetic field, despite having set Stokes profiles with signal $<5\sigma_n$ to zero before the inversion, where $\sigma_n$ is the noise level as determined by the standard deviation in the continuum. Observations of sub-granular, small-scale magnetic loops, with lifetimes of a few minutes and sizes of $1-2\arcsec$, with unambiguous transverse components at the PIL, were made possible with use of an integral field unit such that the spatiotemporal resolution was not compromised. 

Simultaneous observations with visible and near infrared Zeeman diagnostics have also been demonstrated to produce conflicting results, due to a different in formation heights or magnetic substructure \citep{navarro2004,marian2008_vis_NIR}. Indeed, \cite{marian2006} have shown that the different height of formation of the $6301.5/6302.5$~$\mathrm{\AA}$ lines, and their sensitivities to temperature and magnetic field strengths, causes degenerate solutions from inversions of the Stokes vector or difficulty with the line ratio technique. This is predicted to persist even at the  $20$~km resolution of numerical simulations \citep{khomenko2007}.


The unprecedented spatial resolution made possible by the 4-m Daniel K. Inouye Solar Telescope \citep[DKIST;][]{DKIST2020}, combined with the high polarimetric sensitivity and spectral resolution of the Visible SpectroPolarimeter \citep[ViSP;][]{ViSP}, affords the opportunity to observe the fine-structure of the magnetic flux elements in the photosphere. In this study, we investigate the spectropolarimetric properties of small-scale magnetic features of the solar internetwork, as observed by DKIST, with inversions utilised to recover the physical properties from the observed Stokes vector. Studying the small-scale photospheric magnetic field is one of the key research aims of the DKIST Critical Science Plan \citep{dkistcsp}. In Section~\ref{sect:observations} we describe the dataset, its reduction and provide an estimate of the spatial resolution. In Section~\ref{sect:results} we present the analysis of the data, beginning with a description of the inversions. We define three case studies and examine the spectropolarimetric properties and spatial structure of the small-scale magnetic features in detail. Finally we discuss the results in Section~\ref{sect:discussion} before drawing our conclusions in Section~\ref{sect:conclusions}, with a look to future observations.

\section{Observations \& Data Reduction}\label{sect:observations}
\subsection{Observations}
On 26 May 2022 between 17:45 and 19:31 UT observations of a quiet Sun region were acquired with the ViSP at the DKIST during its first cycle of observations from the operations and commissioning phase (OCP1). The observing sequence was designed to obtain narrow, repeated scans of the solar surface at disk centre. The selected slit width was $0{\,}.{\!\!}{\arcsec}041$ with a slit step of $0{\,}.{\!\!}{\arcsec}041$ and $104$ slit positions, giving a map of $4{\,}.{\!\!}{\arcsec}26\ \times 76{\,}.{\!\!}{\arcsec}14$. The cadence between the commencement of a given frame to the start of the next was $13$ minutes and $17$ seconds, and $8$ frames were recorded in total. The first arm of the ViSP captured a spectral region which includes the magnetically sensitive photospheric Fe I line pair at $6301.5$~$\mathrm{\AA}$ and $6302.5$~$\mathrm{\AA}$ (with effective Land\'e g-factors of $1.67$ and $2.5$, respectively), while the third arm captured a spectral region which includes the chromospheric Ca II $8542$~$\mathrm{\AA}$ line. The second arm of the ViSP did not record observations. In this paper we focus our analysis on the Fe I doublet. The spatial sampling along the slit is $0{\,}.{\!\!}{\arcsec}0298$/pixel for the arm which recorded the Fe I $6300$~$\mathrm{\AA}$ spectral region. At each slit step, 24 modulation cycles of 10 modulation states were acquired. The exposure time was $6$~ms, yielding a corresponding $1.44$~seconds total integration time per step. The seeing conditions during the scan were very good, with the adaptive optics (AO) locked during $99.2\%$ of the scan step positions and a mean continuum intensity contrast of $6.2\%$. However, the maximum continuum intensity contrast is closer to $9\%$. The mean Fried parameter reported by the DKIST wavefront correction system was $12$~cm during these observations. The mean standard deviation across all pixels (except those which did not have the AO locked) at a continuum wavelength in Stokes $Q$, $U$, and $V$ in the spectral region adjacent to the Fe I $6302.5$~$\mathrm{\AA}$ line is $7.5\times10^{-4}$~$I_\mathrm{c}$, $7.4\times10^{-4}$~$I_\mathrm{c}$, and $7.5\times10^{-4}$~$I_\mathrm{c}$, respectively. The linear dispersion in the $6300$~$\mathrm{\AA}$ region is $12.85$~$\mathrm{m\AA}$/pixel and the spectral resolution is twice this value.

\begin{figure}
    \centering
    \includegraphics[width=\columnwidth]{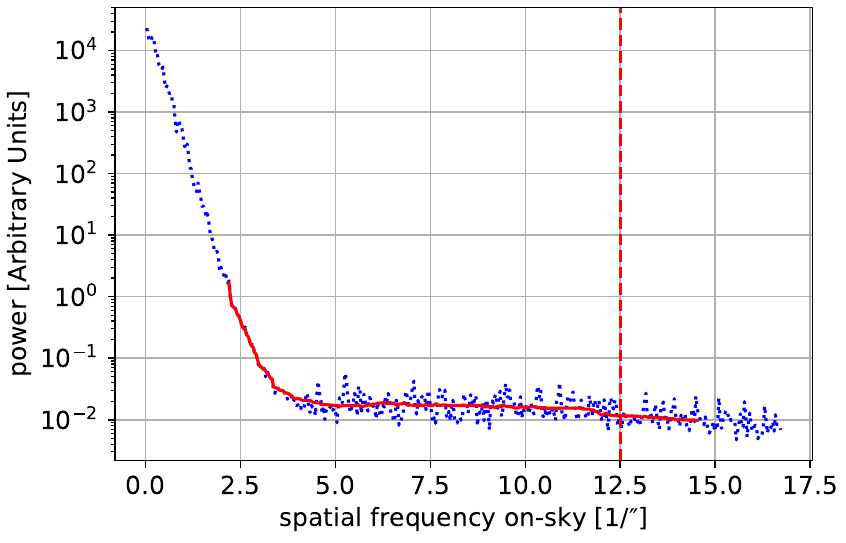}
    \caption{Summed power spectrum of the intensity, computed at a wavelength of $6302.47$~$\mathrm{\AA}$ for each step, summed across step positions with greater than the mean continuum intensity contrast of $6.2\%$. The \textit{dotted, blue line} shows the summed power spectrum  windowed by a Hann filter before computing the FFT. The \textit{solid, red line} shows a median filter applied to the Hann windowed power spectrum. The \textit{dashed, red vertical line} indicates the frequency that equates to a spatial resolution of $0{\,}.{\!\!}{\arcsec}08$.}
    \label{fig:power_spectrum}
\end{figure}

\begin{figure*}
    \centering
    \includegraphics[width=\textwidth]{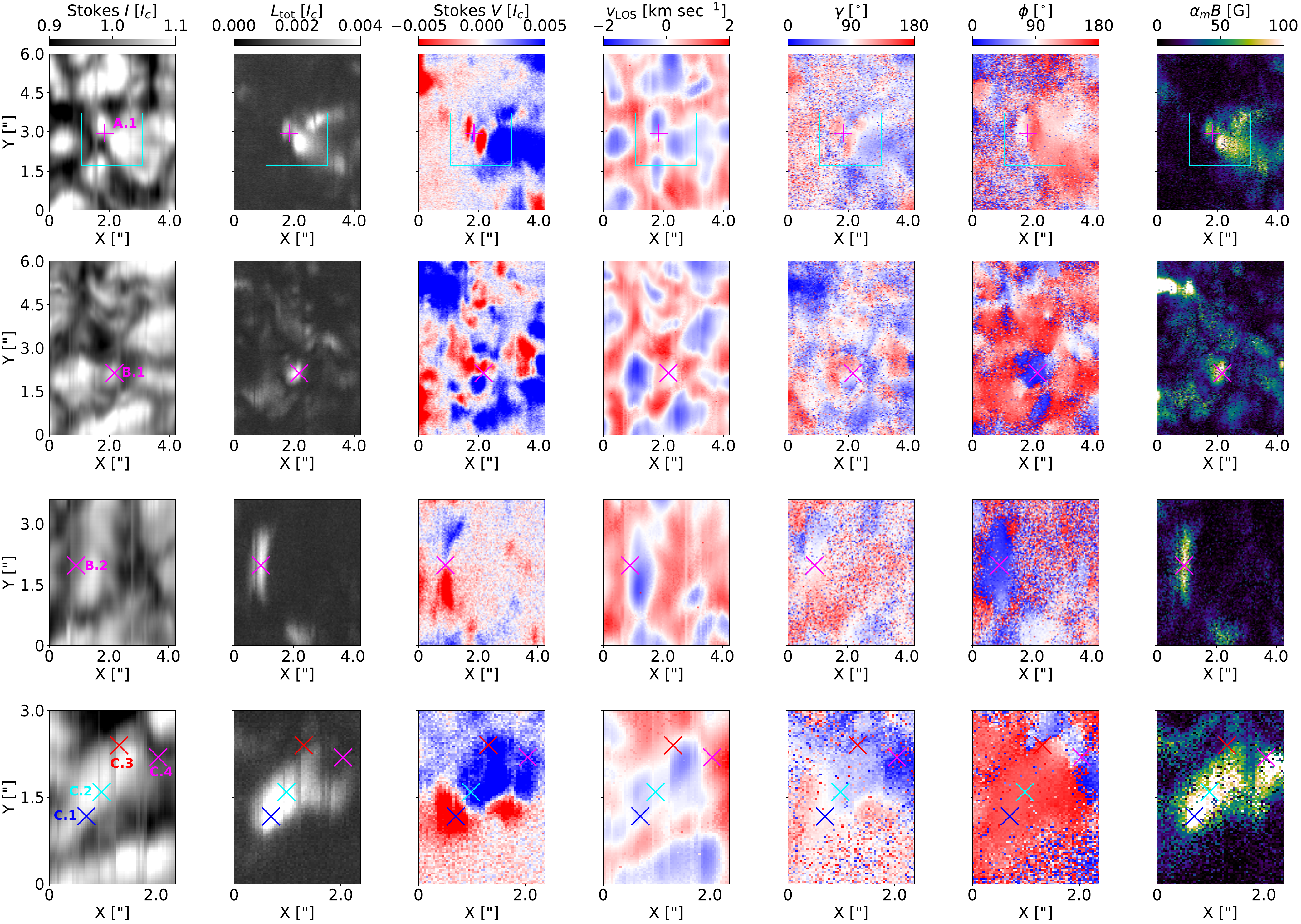}
    \caption{Three case studies of small-scale magnetism, including a serpentine-like magnetic configuration (\textit{upper row}), two magnetic loops (\textit{middle rows}), and a magnetic granule (\textit{bottom row}). From \textit{left to right} is Stokes I at $6303.17$~$\mathrm{\AA}$,  $L_{\mathrm{tot}}$, Stokes $V$ at $6302.45$~$\mathrm{\AA}$, $v_{\mathrm{LOS}}$, $\gamma$, $\phi$, and $\alpha_m B$. The last four parameters are derived from SIR inversions with one magnetic and one non-magnetic model atmosphere (1C configuration). The scanning direction (i.e. solar $X$) is shown on the $x$-axis and the slit direction (i.e. solar $Y$) is shown on the $y$-axis. The labelled (A.1, B.1-B.2, C.1-C.4) markers highlight the spatial location of the sample pixels: the full Stokes vector from pixel A.1 is shown in Fig.~\ref{fig:profileA}, B.1 is shown in Fig.~\ref{fig:profileB1}, B.2 is shown in Fig.~\ref{fig:profileB2}, C.1 is shown in Fig.~\ref{fig:profileC1}, C.2 is shown in Fig.~\ref{fig:profileC2}, and the circular polarisation profiles for C.3 and C.4 are shown in Fig.~\ref{fig:ROIC}. The region outlined by the \textit{cyan box} in the upper row is shown in Fig.~\ref{fig:ROIA}.}
    \label{fig:ROIs}
\end{figure*}

\begin{figure*}
    \centering
    \includegraphics[width=\textwidth]{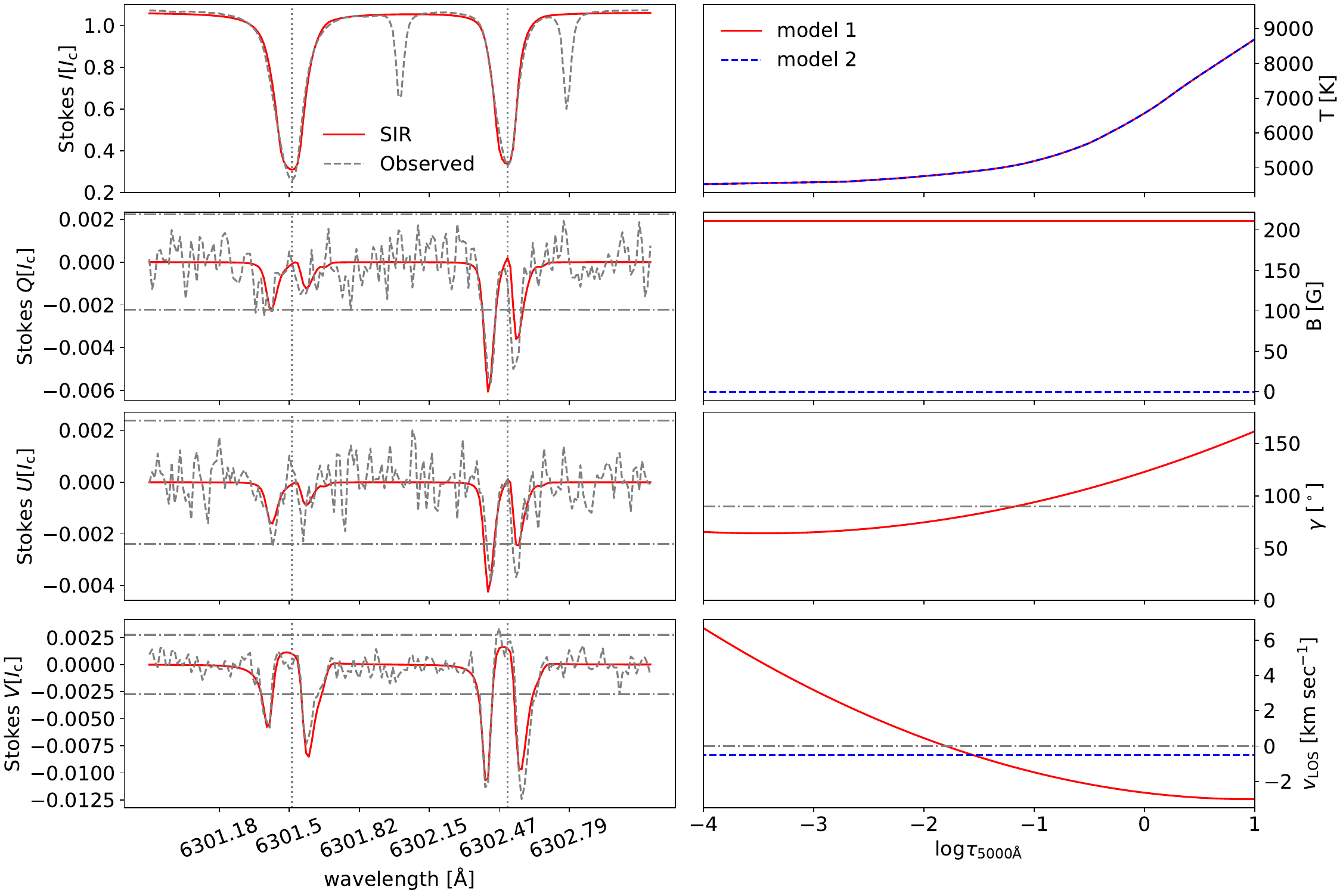}
    \caption{Sample observed Stokes vector with the best-fit inverted profiles (\textit{left panels}) and model associated atmospheres (\textit{right panels}), whose spatial location is labelled in the serpentine magnetic element in the upper row of Fig.~\ref{fig:ROIs} (pixel A.1). The \textit{horizontal} (\textit{dot-dashed}) \textit{lines} show the noise thresholds ($\pm 4\sigma_n$), while the \textit{vertical} (\textit{dotted}) \textit{lines} indicate the rest wavelengths of the Fe I doublet. The retrieved atmospheric parameters are shown as a function of optical depth. The $\phi$, $\alpha_m$, and $v_{\mathrm{mac}}$ values were $107^\circ$, $0.47$, and $0.015$~km~sec$^{-1}$, respectively. This inversion is a 1C configuration.}
    \label{fig:profileA}
\end{figure*}


\begin{figure*}
    \centering
    \includegraphics[width=\textwidth]{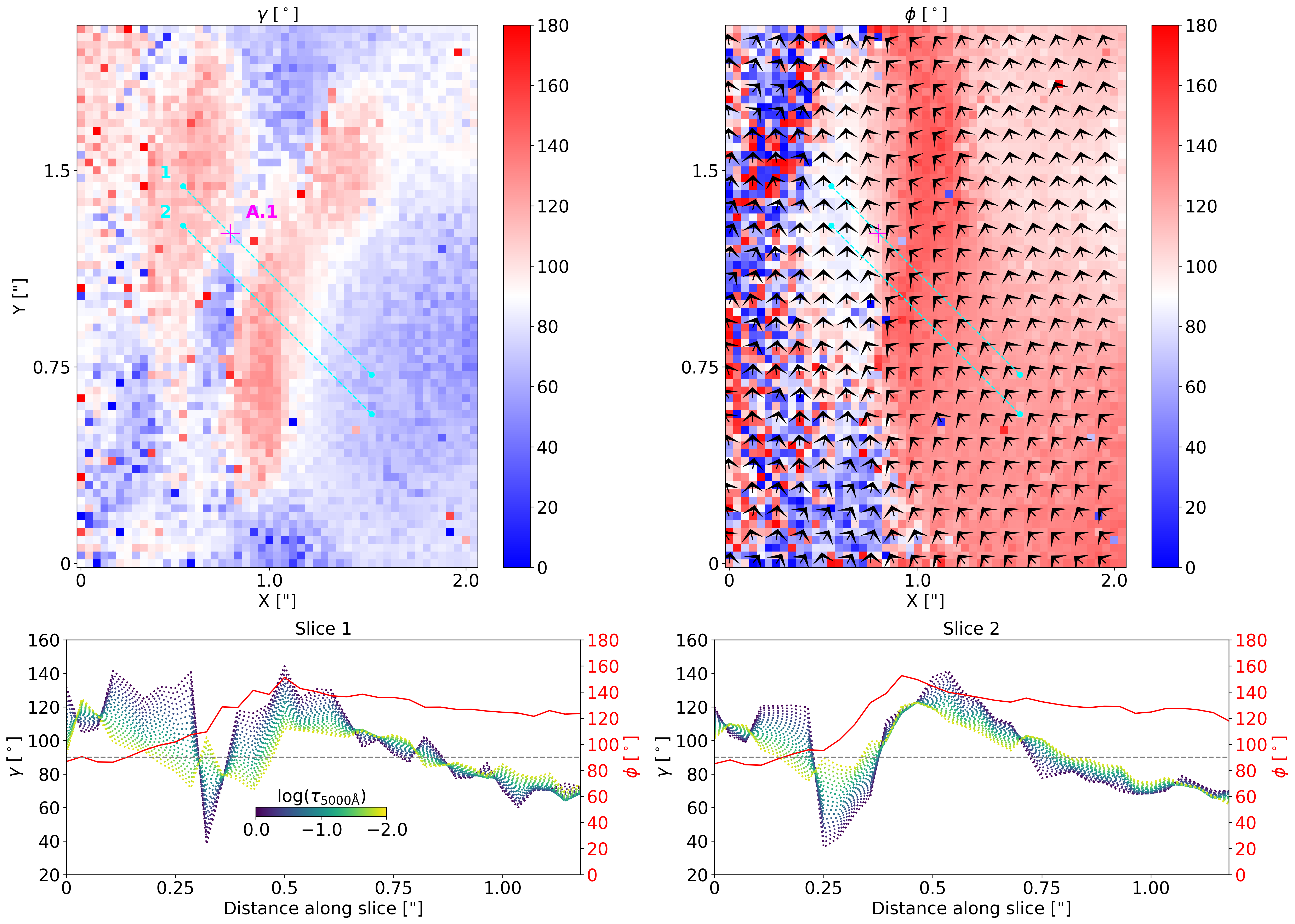}
    \caption{Maps of $\gamma$ (\textit{upper left}) and $\phi$ (\textit{upper right}) in the serpentine structure shown in the first case study of Fig.~\ref{fig:ROIs}. The $\gamma$ map is shown at $\mathrm{log}(\tau_{5000\mathrm{\AA}}) = -1.0$. The direction of the $\phi$ is marked with arrows whose value is given after binning by a $3\times3$ kernel for visual clarity. The azimuth is defined such that a value of 0 is aligned along the line from Solar East to West, increasing anti-clockwise. The variation in $\gamma$ and $\phi$ is also shown across slices labelled $1$ (\textit{lower left}) and $2$ (\textit{lower right}), whose initial and final locations are indicated by the \textit{blue, dashed lines}. The \textit{dotted lines} shows the variation in $\gamma$ at optical depths between $\mathrm{log}(\tau_{5000\mathrm{\AA}}) = -2.0$ and $\mathrm{log}(\tau_{5000\mathrm{\AA}}) = 0.0$, while the \textit{solid, red lines} show the variation in $\phi$, across the slices. The \textit{dashed, gray horizontal lines} mark the $90^\circ$ polarity inversion point for $\gamma$.}
    \label{fig:ROIA}
\end{figure*}

\subsection{Data reduction}
The ViSP calibration and reduction pipeline was applied to the data for dark current removal, flat fielding, and polarimetric calibration \footnote{Version 2.0.2 of the ViSP level 0 to level 1 pipeline was used.}. The polarisation amplitudes in the continuum should be negligible. Stokes $Q$, $U$, or $V$ signal at continuum wavelengths at disk centre can be generated by cross-talk (henceforth referred to as environmental polarisation) with Stokes $I$ when the atmosphere is not ``frozen'' during the modulation process \citep{crosstalk1999}.  We followed the method of \cite{crosstalk1992} to remove environmental polarisation (Stokes $I \rightarrow Q, U, V$ cross-talk only). The two $\mathrm{O}_2$ telluric lines close to the Fe I $6300$~$\mathrm{\AA}$ doublet provide a method for validating that this correction is appropriate, as there should be no residual signal in $Q$, $U$, and $V$ at the wavelengths of these tellurics if the correction is successful. The mean correction amplitude across all pixels in Stokes $Q$, $U$, and $V$, as determined in the continuum adjacent to the Fe I $6302.5$~$\mathrm{\AA}$ line, was $2.8\times10^{-3}$~$I_\mathrm{c}$, $1.8\times10^{-3}$~$I_\mathrm{c}$, and $4.3\times10^{-4}$~$I_\mathrm{c}$, respectively. There remains some small-amplitude artefacts and residual cross-talk in the data, specifically the continuum does not always have net-zero polarisation across the full spectrum. In order to correct this, a simple linear least-squares regression was performed individually to every Stokes $Q$, $U$, and $V$ profile, with the wavelengths of the spectral and telluric lines masked from the function. The linear fit is then subtracted from each profile, resulting in a continuum that is consistently zero at every wavelength, with standard deviations due to photon noise permitted. 

\subsection{Estimating the spatial resolution}
To estimate the maximal spatial resolution in the data, we chose a spectral point in the blue wing of the line at $6302.47$~$\mathrm{\AA}$. For this wavelength, the contrast is increased relative to the continuum. We selected those slit positions for which the contrast is above the mean value of approximately $6.2\%$, and computed the sum of their one-dimensional spatial power at each spatial frequency point. We also applied a Hann filter to each slice before computing the one-dimensional fast fourier transform (FFT), to ensure we do not introduce high spatial frequencies at the edges of each slice. The result is shown in Fig.~\ref{fig:power_spectrum}. Assuming that the noise in the data is additive white Gaussian noise, the frequency at which the spatial power levels can be used as an upper estimate for the noise floor at all spatial frequency points. Increased power above this level clearly indicates the presence of real signal in the data at that spatial frequency. 

We estimate the maximal spatial resolution as $0{\,}.{\!\!}{\arcsec}08$ for this scan. There is significant power at the corresponding spatial frequency even when the Hann window is applied. We also point out that the effective spatial resolution is expected to vary across the scan, perhaps significantly, as a function of the quality of the atmospheric seeing conditions and the performance of the DKIST adaptive optics correcting it. From an inspection of the smallest structures visible in Stokes $I$, we estimate a spatial resolution of at least $0{\,}.{\!\!}{\arcsec}1$ (about $72$~km).

\section{Data analysis $\&$ Results}\label{sect:results}

\subsection{Inversion strategy}
We employed the Stokes Inversion based on Response Functions (SIR) code \citep{SIR} with its Python-based parallelized wrapper \citep{ricardo2021} to invert the full Stokes vector in the Fe I line pair for every pixel in the dataset. The first inversion setup we describe here allows us to produce approximate maps of the full dataset. To achieve this we used a two-component inversion, with one magnetic and one non-magnetic model atmosphere per pixel element, with the contribution of each given by their respective filling factors, $\alpha$. We inverted every pixel $30$ times, with randomised initial values in the magnetic field strength, $B$, line-of-sight (LOS) velocity, $v_{\mathrm{LOS}}$, magnetic inclination, $\gamma$, and magnetic azimuth, $\phi$. The $\gamma$ is defined as the angle between the magnetic vector and the observer's LOS (or solar normal, at disk centre), while the $\phi$ is the angle of the magnetic field vector in the plane perpendicular to the observer's LOS (i.e. in the plane of the solar surface). For each pixel we selected the solution which had the minimum $\chi^2$. Apart from temperature, $T$, which had four nodes, all other parameters, including $B$, had only one node, and thus were forced to be constant in optical depth. In this case, where only one component is magnetic, we explicitly refer to the filling factor of the magnetic component, $\alpha_m$, and the magnetic flux density, $\alpha_m B$. The $T$ in both components was always forced to be the same assuming lateral radiative equilibrium. The microturbulent velocity, $v_{\mathrm{mic}}$ was included as a free parameter independently in each component, and although the macroturbulent velocity, $v_{\mathrm{mac}}$, was also included as a free parameter, it was forced to be the same in both components, assuming it to primarily account for the spectral resolution of the spectrograph. 

We do not explicitly include an unpolarised stray light component to the inversions. Instead, we adopt an inversion with two model atmospheres per pixel element, allowing the filling factor of a non-magnetic model to encapsulate the effect of an unpolarised stray light contribution. Unlike with GREGOR, we do not have an estimate of the unpolarised stray light fraction \citep{borrero2016}. Nevertheless, \cite{Campbell2021b} investigated the statistical impact a varying unpolarised stray light fraction has on the retrieval of $T$, $B$, $v_{\mathrm{LOS}}$, and $\phi$ values using synthetic observations and found only the $T$ contrast was impacted, with the other parameters invariant.

In select cases the inversion will require additional free parameters in order to fit Stokes profiles with asymmetries or abnormal or complex shapes (e.g. a Stokes $V$ profile which does not only have one positive and one negative lobe of equal amplitude). Further, the number of free parameters required may increase significantly when two magnetic components are required to reproduce the observed Stokes vector. We approached these cases with a principle of minimising the number of free parameters as much as possible. In order to reproduce a Stokes vector that is not adequately fit by the initial inversion, we enabled SIR to choose the optimum number of nodes in $B$, $v_{\mathrm{LOS}}$, and $\gamma$ up to a maximum of 10 and up to 3 in $\phi$. We then experimented by systematically reducing the number of nodes in each of these parameters until SIR was unable to fit the profiles with a minimum number of free parameters. In these pixel-specific inversions, we repeated the inversion up to $1000$ times with randomised initial model atmospheres. For clarity, in later sections, we label inversions with two magnetic model atmospheres  as 2-component (2C) inversions. Inversions with one magnetic and one non-magnetic model atmosphere are labelled as 1-component (1C) inversions. In 2C inversions the non-magnetic model atmosphere is replaced by a magnetic one.

Although in forthcoming sections we show the atmospheric parameters at a range of optical depths, we stress that the Fe I $6300$~$\mathrm{\AA}$ doublet is only expected to be sensitive to perturbations in $\gamma$, $\phi$, $B$, and $v_{\mathrm{LOS}}$ in the range between $\mathrm{log}(\tau_{5000\mathrm{\AA}}) = -2.0$ and $\mathrm{log}(\tau_{5000\mathrm{\AA}}) = 0.0$ in a significant way. A detailed analysis of the diagnostic potential of this line, in the context of other photospheric Fe I lines and their response functions, is provided by \cite{cabrera2005} and \cite{quintero2021}.

\subsection{Case studies}
An inspection of the scans revealed an abundance of circular polarisation of both positive and negative polarities, as well as small patches of linear polarisation. It is also clear that there are some particularly strong, large patches of circular polarisation with much weaker magnetic flux patches in their immediate vicinity. In the current work we focus on the weaker, smaller-scale magnetic structures, and curate a few case studies. We used the SIR Explorer (SIRE) tool \citep{SIRE, campbell2023} to locate and analyse these case studies in the full-scan inversions. We define a magnetic loop as two patches of opposite polarity circular polarisation in close contact (i.e. within only a few pixels of each other) with linear polarisation connecting them. This definition indicates that we have two vertical fields (identified by the two opposite polarity circular polarisation signals) which are connected with a horizontal field (identified by the linear polarisation signals) to complete the loop. These structures could be $\Omega-$ or $U-$shaped loops. In the absence of linear polarisation signals, we regard the structure as a bi-pole. In addition to magnetic loops, we also identify unipolar magnetic flux patches with linear polarisation signals but only one polarity of circular polarisation. Finally we define a serpentine structure as one with more than one PIL and clear linear polarisation signals at each PIL. We further classify serpentine magnetic configurations by the number of PILs they have. 

In total we identified $53$ small-scale magnetic elements of interest, including at least $47$ magnetic loops and $4$ unipolar magnetic patches, with linear and circular polarisation detected in all of them. We located several candidate serpentine structures but by examining the azimuth we were able to discern that they were in fact two magnetic loops in close contact due to a sharp discontinuity in azimuth values. In two of the remaining serpentine-like magnetic elements, the polarity of the magnetic field appeared to change three times across the structures, and in one changed only twice. However only one serpentine structure had linear polarisation throughout and at all three PILs. In the following sections we define three case studies from this analysis. We unveil the potential detection of a serpentine magnetic structure in section~\ref{sect:serpentine}, discuss the problem of degenerate solutions at the PILs of magnetic loops in section~\ref{sect:PILs}, and analyse a magnetic granule in section~\ref{sect:granule}. Overview maps of these case studies are shown in Fig.~\ref{fig:ROIs}. The upper row shows the serpentine magnetic structure, the two middle rows show two magnetic loops with transverse magnetism at the PIL, and the bottom row shows the magnetic granule. Defined here is the total linear polarisation,
\begin{equation}
    {L_{\mathrm{tot}}} = \frac{\int_{\lambda_b}^{\lambda_r} [Q^2(\lambda)+U^2(\lambda)]^{\frac{1}{2}}  d\lambda}{I_c \int_{\lambda_b}^{\lambda_r} d\lambda},
\end{equation}
where $\lambda_r$ and $\lambda_b$ are the red ($6302.7$~$\mathrm{\AA}$) and blue ($6302.3$~$\mathrm{\AA}$) limits of integration in wavelength across the $6302.5$~$\mathrm{\AA}$ line, while $I_c$ is the Stokes $I$ signal measured in the neighbouring continuum and spatially averaged along the slit direction for each scan step. Here we also show maps of the azimuth, which we have post-processed to force the values to vary between $0^\circ$ and $180^\circ$. We stress that we have not disambiguated the data.

\begin{figure*}
    \includegraphics[width=\textwidth]{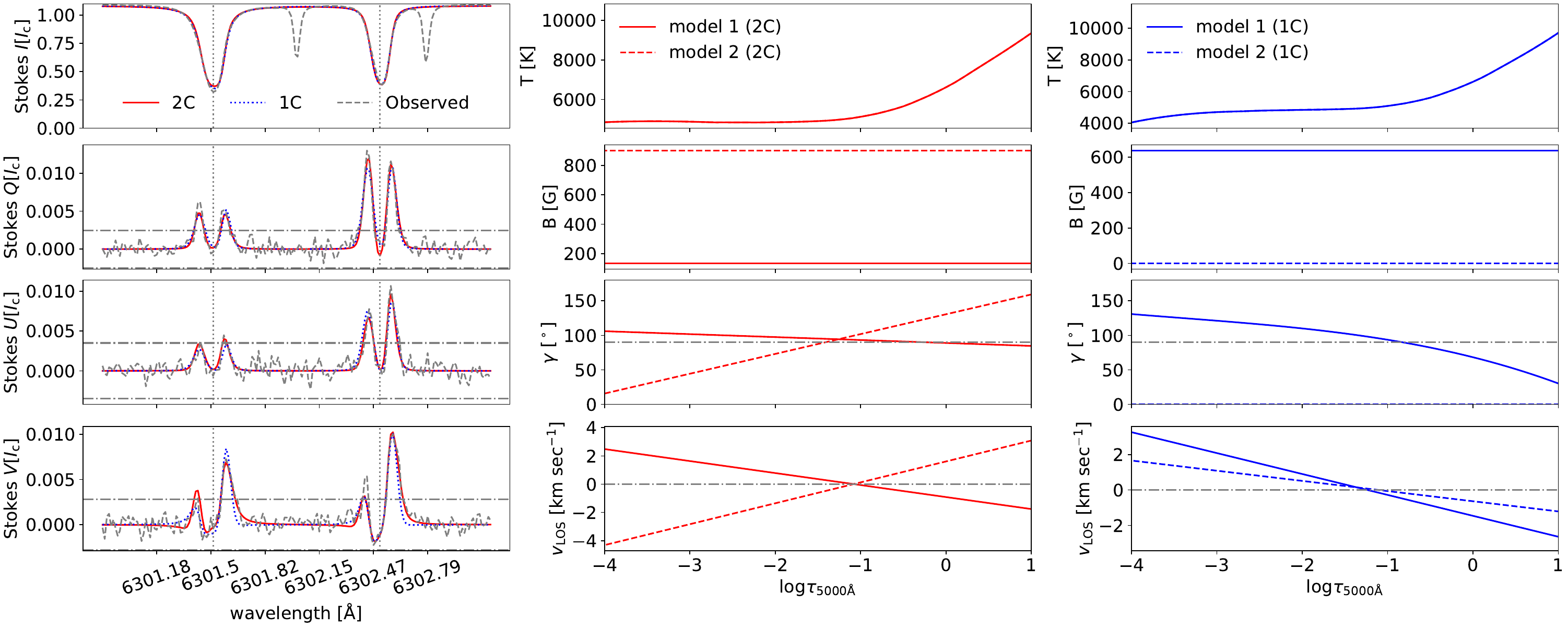}
    \caption{Two solutions to the Stokes vector belonging to pixel B.1 based on two magnetic models (2C) and one magnetic model (1C). The Stokes $V$ signal has a complex shape. In the \textit{left column} the \textit{red, solid lines} show the synthetic Stokes vector for 2C, and the \textit{blue, dotted lines} show the same for 1C. The model parameters for 2C and 1C are shown in the \textit{middle column and right column}, respectively. The location of this pixel is indicated in Fig.~\ref{fig:ROIs}. For 2C, the $\alpha$ values of model 1 and 2 were $0.93$ and $0.07$, respectively, the $\phi$ values were $27^\circ$ and $191^\circ$, respectively, and the $v_{\mathrm{mac}}$ was $0.77$~km~sec$^{-1}$ for both models. For 1C, the $\alpha$ values of model 1 and 2 were $0.17$ and $0.83$, respectively, the $\phi$ value was $22^\circ$, and the $v_{\mathrm{mac}}$ was $0.94$~km~sec$^{-1}$ for both models.}
    \label{fig:profileB1}
\end{figure*}

\begin{figure*}
    \centering
    \includegraphics[width=\textwidth]{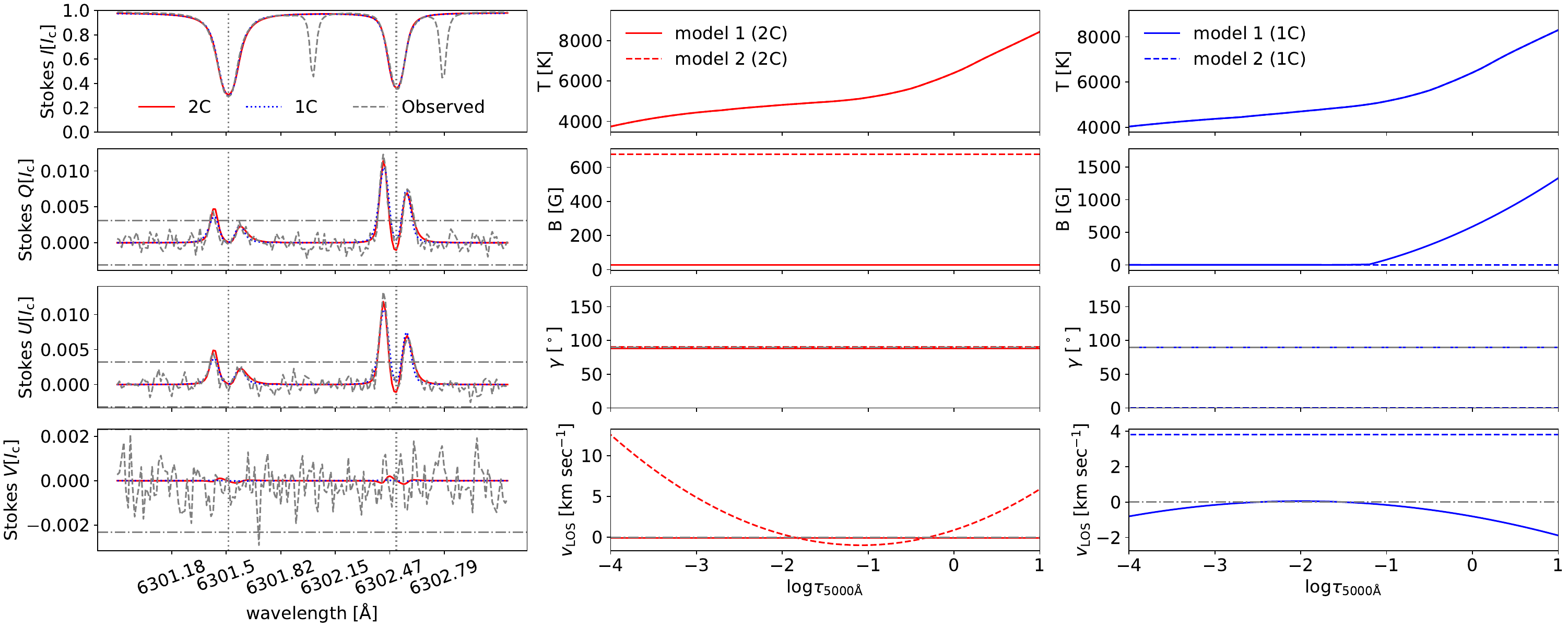}
    \caption{As in Fig.~\ref{fig:profileB1}, but for the Stokes vector belonging to pixel B.2. For 2C, the $\alpha$ values of model 1 and 2 were $0.83$ and $0.17$, respectively, the $\phi$ values were $158^\circ$ and $23^\circ$, respectively, and the $v_{\mathrm{mac}}$ was $0.71$~km~sec$^{-1}$ for both models. For 1C, the $\alpha$ values of model 1 and 2 were $0.93$ and $0.07$, respectively, the $\gamma$ and $\phi$ values for model 1 were $90^\circ$ and $203^\circ$, respectively and the $v_{\mathrm{mac}}$ was $0.94$~km~sec$^{-1}$ for both models. There is no significant Stokes $V$ signal.}
    \label{fig:profileB2}
\end{figure*}

\subsubsection{Case study I: Small-scale serpentine magnetism}\label{sect:serpentine}
 
The first case study is shown in the upper row of Fig.~\ref{fig:ROIs}. Upon initial examination at a wavelength of $6302.39$~$\mathrm{\AA}$, this appeared to be a magnetic loop, where two patches of circular polarisation are in close contact and linear polarisation bridges the two patches. Upon closer inspection, a small adjustment in the wavelength position in SIRE reveals a circular polarisation pattern that is more complex than a single magnetic loop. When inspected at a wavelength of $6302.45$~$\mathrm{\AA}$, closer to the rest wavelength of the line, as shown in Fig.~\ref{fig:ROIs}, the complexity of the magnetic topology in this structure became apparent.

If one examines the $\gamma$ map, it is clear that the polarity of the magnetic field changes sign at least three times across the structure (as opposed to once, as would be expected in a simple U- or $\Omega$-shaped loop). When the pixels in this structure have reasonably symmetrical, two-lobed Stokes $V$ profiles, and linear polarisation is abundant, the change in $\gamma$ is unambiguous and SIR produces good fits to the observed Stokes vectors. However, there is a narrow location at which complex Stokes $V$ profiles are found and the initial inversion is unable to fit these profiles. A sample Stokes vector from this location, at the edge of the granule, whose spatial location is highlighted by the marker in Fig.~\ref{fig:ROIs}, is shown in Fig.~\ref{fig:profileA}. This Stokes vector has all three polarisation parameters with signal greater than the $4\sigma_n$ level, but the Stokes $V$ profile has two negative lobes. We therefore increased the number of free parameters permitted in $B$, $\gamma$, and $v_{\mathrm{LOS}}$ to $10$ each, and in $\phi$ to $3$, and repeated the inversion with $200$ randomised initialisations per pixel, and systemically reduced the number of nodes in each parameter. We found that the best fits could be achieved with $1$ node in $B$, but $3$ each in $\gamma$ and $v_{\mathrm{LOS}}$ for the magnetic component. For the non-magnetic component we found only $2$ nodes in $v_{\mathrm{LOS}}$ were required. The synthetic vector shown in Fig.~\ref{fig:profileA} resulted from this process. We then ran the inversion with this configuration in a small area, shown in Fig.~\ref{fig:ROIA}, so we could investigate how $\gamma$ changed as a function of optical depth. 

Shown in Fig.~\ref{fig:ROIA} is the variation in $\gamma$ across two $1{\,}.{\!\!}{\arcsec}22$ ($882$~km) slices generated by cubic interpolation between two points, as a function of optical depth between $\mathrm{log}(\tau_{5000\mathrm{\AA}}) = -2.0$ and $\mathrm{log}(\tau_{5000\mathrm{\AA}}) = 0.0$, where the spectral line is expected to be responsive to changes in $\gamma$ according to the response functions. Regardless of the optical depth considered, the polarity of the magnetic field changes three times across the slices. However, the spatial position at which this polarity inversion occurs can depend on the optical depth. We emphasise that this is occurring on a small, sub-granular scale, as from the point where the first polarity inversion occurs to the last is a distance of about $400$~km in slice $1$.

To truly characterise this magnetic element as serpentine we must also examine the azimuth. If the azimuth showed a sharp, discontinuous change between two distinct values this could indicate that we have actually observed two separate magnetic loops that are not connected. Figure~\ref{fig:ROIA} also shows the variation in $\phi$ across the slices. In the upper (lower) slice, $\phi$ varies smoothly from a minimum of $86^\circ$ ($84^\circ$) to a maximum of $151^\circ$ ($153^\circ$). To confidently constrain the azimuth one needs to measure both Stokes $Q$ and $U$. However, the amplitude of the noise in the weaker linear polarisation parameter places a limit on the value that $\phi$ could take when only one is confidently measured. Therefore, in these cases, and especially when the amplitude of the stronger linear polarisation parameter is large, the $\phi$ value can still be constrained. When heavily noise contaminated, maps of $\phi$ appear random. The map of $\phi$ in Fig.~\ref{fig:ROIA} is smooth-varying across the magnetic element, including in those pixels where only one linear polarisation parameter has a maximum amplitude greater than $4\sigma_n$. This is in contrast to areas surrounding the magnetic element, where the map of $\phi$ has a salt-and-pepper pattern. The spatial coherence of this map indicates that there is enough linear polarisation signal available in the magnetic element to constrain the azimuth and deduce that it varies across the slices. While a change of $69^\circ$ in $\phi$ is not an insignificant variation, we note that the change occurs gradually. This suggests the magnetic field vector is changing direction not just along the LOS, but in the plane of the solar surface across the slices, but not in a sharp, discontinuous way that would indicate the existence of two distinct magnetic elements. Indeed, after the $\phi$ value reaches its peak in each slice, it continues to vary gradually.

We also repeated the inversion in the small area in Fig.~\ref{fig:ROIA} where the maximum number of nodes in $B$ in the magnetic component was increased from $1$ to $3$, and the maximum number of nodes in $v_{\mathrm{LOS}}$ in the non-magnetic component was increased from $2$ to $3$. We found that the increased number of free parameters made a very marginal improvement in the quality of the fits in a small number of pixels, but made no significant difference to the $\gamma$ or $\phi$ variations in this magnetic element; most importantly, the polarity of the magnetic field changed three times across both slices. We found no evidence that SIR made use of the additional free parameter in $v_{\mathrm{LOS}}$.

\subsubsection{Case study II: Degenerate solutions at polarity inversion lines}\label{sect:PILs}

\begin{figure*}
    \includegraphics[width=\textwidth]{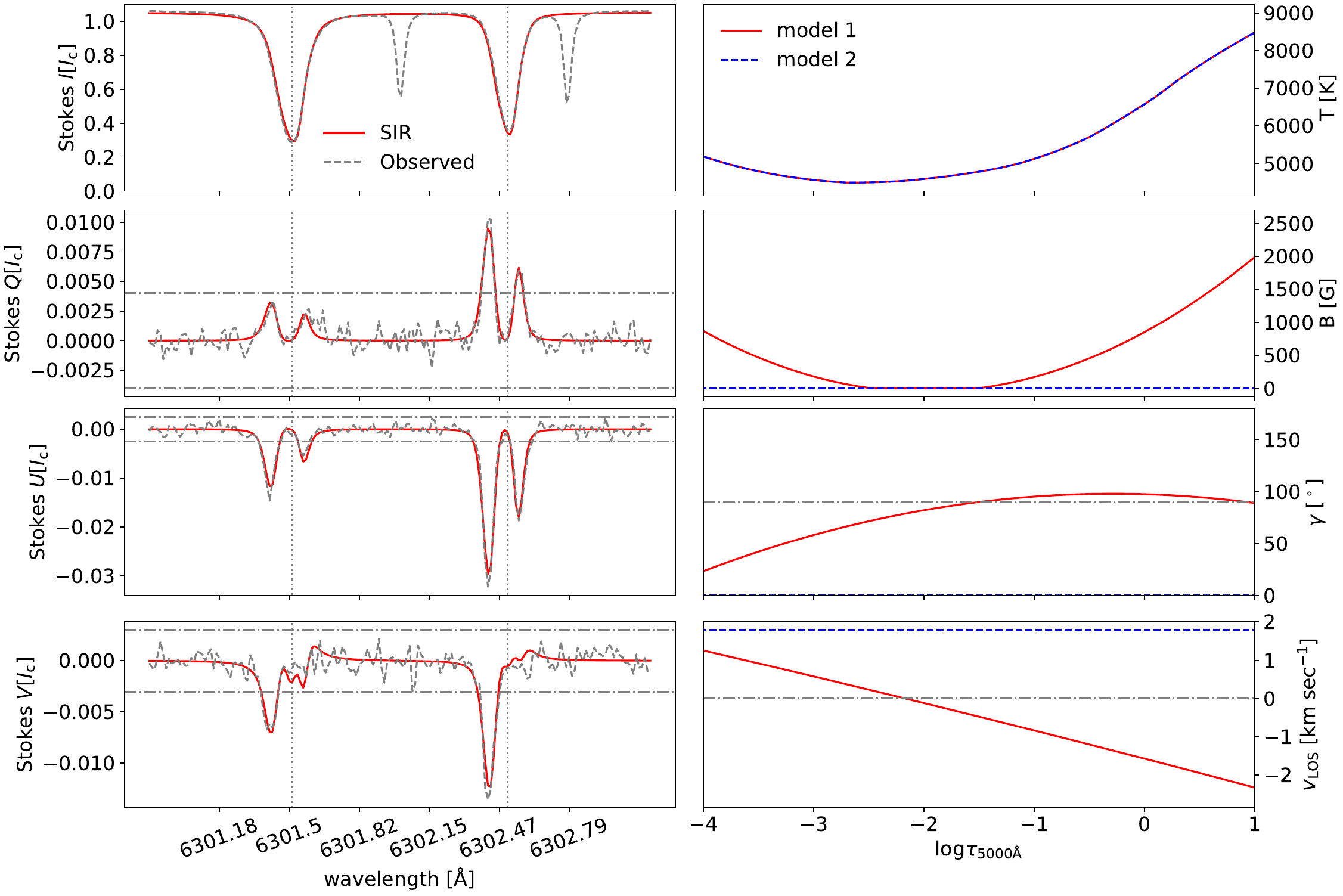}
    \caption{As in Fig.~\ref{fig:profileA}, but for vector C.1. The filling factors of model 1 and 2 were $0.79$ and $0.21$, respectively, while the $\phi$ value was $146^\circ$. The $v_{\mathrm{mac}}$ was $0.73$~km~sec$^{-1}$ for both models. This inversion is a 1C configuration. The Stokes $V$ signal has a single lobe.}
    \label{fig:profileC1}
\end{figure*}

\begin{figure*}
    \centering
    \includegraphics[width=\textwidth]{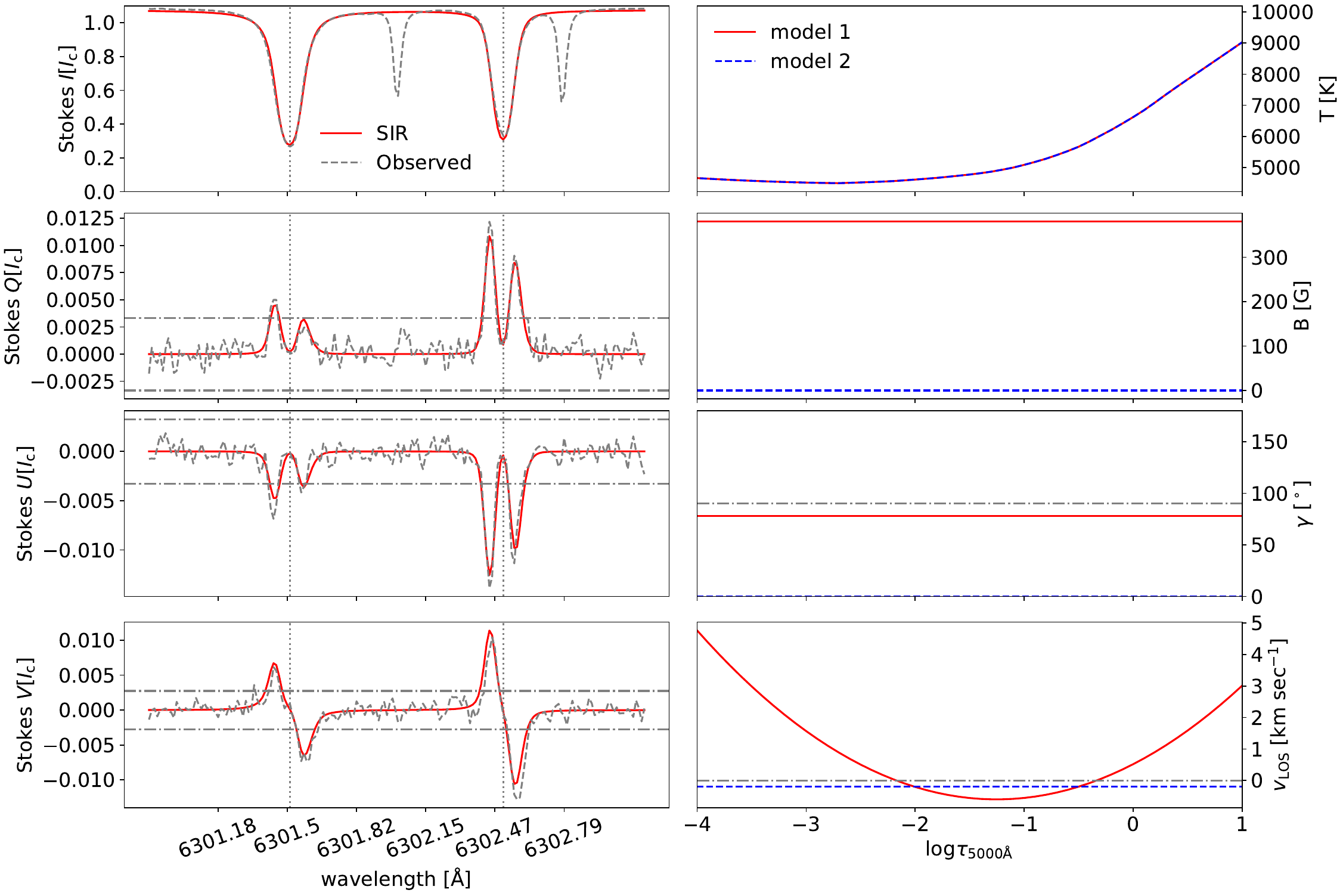}
    \caption{As in Fig.~\ref{fig:profileA}, but for vector C.2. The filling factors of model 1 and 2 were $0.31$ and $0.69$, respectively, while the $\gamma$ and $\phi$ values were $78^\circ$ and $152^\circ$. The $v_{\mathrm{mac}}$ was $0.95$~km~sec$^{-1}$ for both models. This inversion is a 1C configuration. The Stokes $V$ signal exhibits a regular double-lobed, anti-symmetric shape.}
    \label{fig:profileC2}
\end{figure*}

\begin{figure*}
    \centering
    \includegraphics[width=\textwidth]{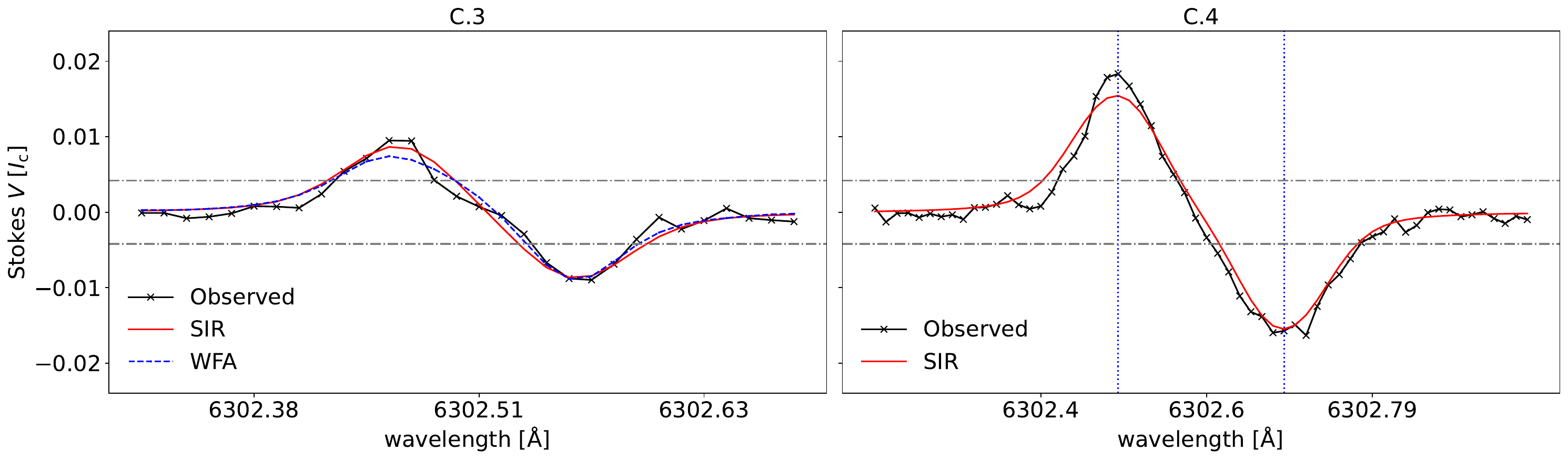}
    \caption{In the \textit{lower panels} the Stokes \textit{V} profiles from pixels C.3 and C.4 are shown on the \textit{left} and the \textit{right}, respectively. Profile C.3 has over-plotted the WFA fit obtained from the derivative of Stokes $I$ (\textit{blue, dashed line}). Both profiles have the synthetic SIR profile over-plotted (\textit{red, solid line}). The \textit{horizontal, grey, dashed-dotted lines} show the $4\sigma_n$ threshold and the \textit{vertical, blue, dotted lines} show the locations of the blue and red lobes used for the SFA estimate for $B$.}
    \label{fig:ROIC}
\end{figure*}

One of the most likely places to find complex, abnormal Stokes profiles (i.e. a larger or smaller number of lobes than expected, or with significant asymmetries, or both) in the quiet Sun is at, or near, the PIL in magnetic loops. In the middle rows of Fig.~\ref{fig:ROIs} we group together two simple magnetic loops for the second case study. These structures have in common that linear polarisation is found at the PIL, however, they differ significantly in terms of their circular polarisation signals at the same location. The magnetic loop shown in the second row in Fig.~\ref{fig:ROIs} is found in the intergranular lane (IGL) next to a tiny granule. From the azimuth map, we can immediately determine that this magnetic loop is a distinct structure compared to its surroundings. The sample Stokes vector B.1 shown in Fig.~\ref{fig:profileB1} demonstrates that the linear polarisation signals are strong, with small asymmetries, while the circular polarisation signal has a three-lobed shape, and SIR is unable to reproduce it without an increase in the number of free parameters. The middle panel in Fig.~\ref{fig:profileB1} shows the atmospheric parameters which resulted in a good fit to the observed Stokes vector with two magnetic model atmospheres (2C), including $1$ node in $B$, $2$ in $v_{\mathrm{LOS}}$, $2$ in $\gamma$, and $1$ in $\phi$ per model. SIR has selected one magnetic component with a larger filling factor ($\alpha = 0.93$) that is weaker ($B = 133$~G) to co-exist with another magnetic component with a much smaller filling factor ($\alpha = 0.07$) and much stronger magnetic field ($B = 902$~G). The superposition of the opposing linear stratifications in $v_{\mathrm{LOS}}$ and $\gamma$ is essential to produce the synthetic vector, in particular to produce the circular polarisation profile which could result both from the mixing of opposite polarity Stokes $V$ signals within the spatiotemporal resolution element and from gradients in $v_{\mathrm{LOS}}$. Reducing the number of nodes in either of these parameters significantly degraded the fit.

This sample pixel is not unique in this magnetic loop, as many similar Stokes $V$ profiles are found along the PIL. Furthermore, the solution provided by the inversion is not unique, and the mixing of two magnetic components in the spatiotemporal resolution element is not the only physical scenario that could be responsible for this Stokes vector. We encounter a degeneracy problem, where very different physical solutions provide equally plausible fits to the observed Stokes vector, when trying to fit complex profiles. In Fig.~\ref{fig:profileB1} we also present an alternative, degenerate solution that does not require a 2C configuration. In this case, a 1C configuration with a $638$~G magnetic field that is constant in optical depth was sufficient, but both the magnetic and non-magnetic model atmospheres required strong linear gradients in $v_{\mathrm{LOS}}$ achieved with $2$ nodes each, and crucially the magnetic model atmosphere required a polynomial stratification in $\gamma$ that was achieved with $5$ nodes.

The second magnetic loop shown in the third row in Fig.~\ref{fig:ROIs} is a narrow patch of magnetic flux along the granule-IGL boundary. The structure is about $1{\,}.{\!\!}{\arcsec}5$ long but only $0{\,}.{\!\!}{\arcsec}5$ wide. The Stokes vector from pixel B.2 is taken from the PIL, but in contrast to the former magnetic loop (pixel B.1) there is a complete absence of a Stokes $V$ signal. However, despite this, we once again encounter degenerate solutions as there are large asymmetries in the amplitudes of the $\sigma$-lobes of the linear polarisation profiles. Figure~\ref{fig:profileB2} show solutions based on 1C and 2C configurations. These solutions have in common that SIR selected an almost perfectly transverse magnetic field. Indeed, as we have no circular polarisation we also find there is no benefit to including a gradient in $\gamma$. In the 1C solution, we required $3$ nodes each in $B$ and $v_{\mathrm{LOS}}$ for the magnetic model atmosphere. The $\alpha_m$ for this solution was very large ($\alpha_m = 0.93$) and the magnetic model atmosphere is nearly stationary at optical depths where the response functions peak, while the non-magnetic model atmosphere has a strong down-flow ($v_{\mathrm{LOS}} = 3.8$~km~sec$^{-1}$). We find that the stratification in $B$, decreasing with height until reaching zero near where the response functions peak, is essential. Additionally, a linear gradient in $v_{\mathrm{LOS}}$ was insufficient to fully achieve the asymmetric amplitude of Stokes $Q$ and $U$. As for the 2C solution, we found it to be essential that the model atmospheres have different $v_{\mathrm{LOS}}$ stratifications and different $\phi$ values. The primary model atmosphere, with the larger filling factor ($\alpha = 0.83$), has a very weak magnetic field strength ($B = 28.0$~G) and is nearly stationary in the LOS ($v_{\mathrm{LOS}}$ = -0.1~km~sec$^{-1}$), while the secondary model atmosphere has a polynomial stratification in $v_{\mathrm{LOS}}$ and has a much larger magnetic field strength ($B = 676$~G). We were able to reduce the number of nodes in $B$, $\gamma$, $\phi$, and $v_{\mathrm{LOS}}$ to 1 in both model atmospheres with the exception of the $v_{\mathrm{LOS}}$ of the secondary model atmosphere, which had $3$ nodes and could not be reduced without losing the required asymmetries in the linear polarisation profiles.

We tested the inversions for pixels B.1 and B.1 under 1C and 2C configurations where we allowed the temperature of both model atmospheres to vary independently. In all cases the increase in the number of free parameters did not improve the quality of fit, and further the key features of the parameter stratifications in $B$, $\gamma$, and $v_{\mathrm{LOS}}$ were invariant. Ultimately we found no evidence that increasing the number of free parameters provided either a significantly different solution or an improvement to the quality of the fit to the observed Stokes vector, so we favour the simpler solutions.

\subsubsection{Case study III: Magnetic anatomy of a granule}\label{sect:granule}

The bottom row in Fig.~\ref{fig:ROIs} shows our final case study. Notably, we find there is significant linear and circular polarisation signal throughout almost the entirety of the granule. We take this opportunity to explore the different spectropolarimetric and magnetic properties of the granule and surrounding IGLs, which is the convective building block of the solar surface, for the first time at $70$~km resolution. To this end we select four representative pixels. To begin, in the lower, left-most patch of circular polarisation, linear polarisation is pervasive, while both the linear and circular polarisation profiles are highly asymmetric and the circular polarisation is even single-lobed. Fig.~\ref{fig:profileC1} shows a sample Stokes vector (pixel C.1) from this patch, with its location at the granule-IGL boundary indicated by the blue marker in Fig.~\ref{fig:ROIs}. The initial 1C inversion was unable to accurately fit the asymmetric Stokes profiles, and so the number of free parameters was increased. The very good fit shown in Fig.~\ref{fig:profileC1} is the result, which was achieved with $3$ nodes in $B$, $3$ in $v_{\mathrm{LOS}}$, $3$ in $\gamma$, and $1$ in $\phi$ for the magnetic model atmosphere, and only $1$ node in $v_{\mathrm{LOS}}$ for the non-magnetic model atmosphere. We found that a 1C configuration was sufficient and did not require a 2C configuration. The $B$ stratification required to fit this profile is a gradient which decreases with increased height in the atmosphere, from very strong, kG magnetic field with $B > 2000$~G in the deepest layers. However, we stress that SIR is simply extrapolating to $\mathrm{log}(\tau_{5000\mathrm{\AA}}) = 1.0$. At $\mathrm{log}(\tau_{5000\mathrm{\AA}}) = 0.0$ the field strength is $851$~G. The magnetic model atmosphere ($\alpha_m = 0.79$) is highly transverse in the optical depth range the lines are responsive to but has a gradient which changes polarity and becomes more vertical higher in the atmosphere. In order to produce the asymmetries in all four Stokes parameters, the magnetic model atmosphere also required a strong linear gradient in $v_{\mathrm{LOS}}$ while the non-magnetic component has a single, positive $v_{\mathrm{LOS}}$ value throughout the atmosphere.

Deep in the granule, in the central patch of circular polarisation, linear polarisation is again pervasive but the circular polarisation is much more symmetric and two-lobed. Fig.~\ref{fig:profileC2} shows a sample Stokes vector (pixel C.2) whose location is indicated by the cyan marker in Fig.~\ref{fig:ROIs}. The synthetic vector provided by the initial 1C inversion is a good fit in this case, with the exception of small asymmetries in the linear polarisation profiles. By adding a gradient to the $v_{\mathrm{LOS}}$ of the magnetic model atmosphere, SIR was able to obtain better fits. The $B$ of this pixel is moderate but the pixel had a large $\alpha_m$ such that the $\alpha_m B$ is relatively large ($B = 380$~G, $\alpha_m = 0.31$, $\alpha_m B = 118$~Mx~cm$^{-2}$). In the lower, right patch of circular polarisation, most of the profiles do not have significant linear polarisation and the circular polarisation profiles are typically asymmetric. While the profiles may be asymmetric or not, the change in polarity is nevertheless unambiguous between these three patches of small-scale magnetism. It is perhaps not surprising that the profiles are asymmetric, and therefore indicative of velocity gradients, in pixel C.1 given that it is located at the granule-IGL boundary. Pixel C.2, on the other hand, is more firmly located in the granule. 

The Stokes vectors found in the IGLs surrounding the examined granule were also inspected. Stokes $V$ profiles with varying degrees of asymmetries are found. We find that the field strengths at these locations differ by two orders of magnitude, and at the high end of these extremes the $B$ exceeds $2$~kG. We are motivated to sanity check these values with the weak field approximation (WFA) and strong field approximation (SFA) in order to ensure the inversions are well calibrated \citep{Campbell2021a,campbell2023}, by estimating the longitudinal magnetic flux density with the WFA and magnetic field strength with the SFA. 

In the weak regime, as \cite{pol} demonstrate, the amplitude of Stokes $V$ is proportional to the longitudinal magnetic flux density (where $B_\parallel = B\cos\gamma$),
\begin{equation}\label{eqn:WFA}
    V(\lambda) \approx - \Delta\lambda_B \cos\gamma \frac{dI(\lambda)}{d\lambda},
\end{equation}
where, 
\begin{equation}\label{eqn:SFA}
\lambda_B \approx 4.6686\times10^{-13} \lambda_0^2 g_{\mathrm{eff}}B,
\end{equation}
where $\lambda_0$ is the rest wavelength and $g_{\mathrm{eff}}$ is the effective Land\'e g-factor of the spectral line. Equation~\ref{eqn:WFA} is only valid if the magnetic Zeeman splitting is negligible relative to the Doppler width of a line (i.e. $\Delta\lambda_B/\Delta\lambda_d \ll 1$) and when $\gamma$, $\phi$, $B$, $v_{\mathrm{LOS}}$, the Doppler width, and any broadening mechanisms are invariant along the LOS in the formation region of the line. On the other hand, by measuring $2\Delta\lambda_B$ as the separation of the lobes of Stokes $V$ when the field is strong enough and when $\Delta\lambda_B/\Delta\lambda_d \gg 1$, it is possible to obtain $B$ directly through Eqn.~\ref{eqn:SFA} \citep{khomenko2003, Nelson2021, campbell2023}. As in \cite{ramos2011}, writing the expressions for $\Delta\lambda_B$ and $\Delta\lambda_d$ allows us to say a line is in the weak field regime when the magnetic field strength fulfills:
\begin{equation}\label{eqn:WFA_limit}
    B \ll \frac{4\pi m c}{g_{\text{eff}}\lambda_0 e}\sqrt{\frac{2kT}{M} + v_{mic}},
\end{equation}
where $m$ is the mass of an electron, $c$ is the speed of light, $\lambda_0$ is the rest wavelength of the spectral line, $e$ is the charge of an electron, $k$ is the Boltzmann constant, and $M$ is the mass of the species. For the $6302.5$~$\mathrm{\AA}$ line, assuming $v_{\text{mic}} = 1$~km~sec$^{-1}$ and $T = 5800$ K, we estimate a limit of  $760$~G. Raising the $T$ to $7000$~K would increase the limit to $809$~G. Eliminating the $v_{\mathrm{mic}}$ would decrease the limit to $610$~G.

We sampled pixels at the upper-most granular boundary with low-amplitude, symmetrical Stokes $V$ profiles, and applied the WFA. The circular polarisation profile from the pixel labelled C.3 in Fig.~\ref{fig:ROIs} is shown in Fig~\ref{fig:ROIC}. In the initial 1C inversion, SIR fit this profile with a model atmosphere with $B = 73.4$~G, $\gamma = 57.0^\circ$, $\alpha_m = 0.5$, and $v_{\mathrm{LOS}} = 0.1$~km/sec. Therefore, according to SIR, $\alpha_m B \cos \gamma = 21.2$~Mx~cm$^{-2}$. The $B$ value is an order of magnitude smaller than the limit we estimated using Eqn.~\ref{eqn:WFA_limit}, so we are satisfied that using the WFA on this pixel is appropriate. Meanwhile,  the WFA estimate is obtained from the derivative of Stokes $I$ and through Eqn.~\ref{eqn:WFA}  as  $20.0$~Mx~cm$^{-2}$. The pixel with the weakest $B$ that we found which satisfied the $4\sigma_n$ amplitude threshold, and which had symmetrical lobes, was $25$~G, or $\alpha_m B \cos \gamma = 12.8$~Mx~cm$^{-2}$. For the same pixel, the WFA estimate was $11.7$~Mx~cm$^{-2}$.

We sampled pixels in the right-most IGL which has an extremely strong magnetic field and applied the SFA by measuring the separation of the lobes of Stokes $V$. The circular polarisation profile from the pixel labelled C.4 in Fig.~\ref{fig:ROIs} is shown in Fig.~\ref{fig:ROIC}. In this case, the initial 1C inversion fit this profile with a very strong magnetic field ($B = 2048$~G, $\alpha_m = 0.07$, $\alpha_m B = 139.3$~Mx~cm$^{-2}$) and the positive and negative lobes of Stokes $V$ are separated by $15$ increments in wavelength. The SFA estimate is thus $2078$~G as a single increment in wavelength is equivalent to $138.55$~G. We are both satisfied that using the SFA is appropriate and that the magnetic field strength is accurate as there are many pixels in this IGL with field strengths greater than $1.8$~kG. 

\section{Discussion}\label{sect:discussion}


We have presented the first spectropolarimetric observations of the quiet Sun with DKIST, the first $4-$m class and largest solar optical telescope ever built. With an estimated spatial resolution of $0{\,}.{\!\!}{\arcsec}08$, these observations represent the highest spatial resolution full-vector spectropolarimetric observations ever obtained of the quiet Sun; previous high resolution observations were achieved by the Swedish Solar Telescope \citep[$0{\,}.{\!\!}{\arcsec}16$, e.g.][]{sst2011}{}{}, the Sunrise balloon-borne experiment \citep[$0{\,}.{\!\!}{\arcsec}14-0{\,}.{\!\!}{\arcsec}16$, e.g.][]{lagg2010,wiegelmann2013}{}{}, the GREGOR telescope \citep[$0{\,}.{\!\!}{\arcsec}3-0{\,}.{\!\!}{\arcsec}4$, e.g.][]{lagg2016,campbell2023}{}{}, and the Hinode spacecraft \citep[$0{\,}.{\!\!}{\arcsec}32$, e.g.][]{lites2008,rubio2012}{}{}. After locating at least $53$ magnetic elements of interest, we curated three case studies and examined their physical properties in detail to showcase how the small-scale magnetic fields are spatially organised at this resolution. 

We examined a particular magnetic element where the polarity inverts three times across the structure (see Fig~\ref{fig:ROIA}). In addition to a complex topology in terms of the inclination angle, we also observe the azimuthal angle changing significantly across its structure. We believe that the evidence indicates that it is plausible that the magnetic field lines of this small-scale magnetic structure are serpentine with significant, coherent variation in the plane of the solar surface. We are not aware of any previous studies that have observed these complex magnetic structures in the quiet Sun on a sub-granular scale. Hinode observations have demonstrated that in simple magnetic loops the azimuth can vary significantly in a few minutes \citep{marian2010} and simulations of the twisting of magnetic field lines during small-scale reconnection events show that gradual azimuthal changes in the photosphere are plausible in more complex magnetic topologies \citep{danilovic2009,hansteen2017}. Perhaps it is possible that these complex topologies have not been observed at facilities with lower diffraction limited resolutions than DKIST precisely because they lack the required angular and spectral resolution. Similarly, the complex nature of the serpentine feature could be lost with poorer seeing conditions. However, we also emphasise that this is the only example of serpentine magnetism that we discovered in the dataset which had three PILs and linear polarisation signals throughout the structure, and in the other magnetic loops we see a much more simple variation in $\gamma$, and typically insignificant variation in $\phi$.

An alternative interpretation would be that we observed two magnetic loops in very close proximity. The fact that the azimuth increases to reach a maximum in one part of the slice but then decreases somewhat less significantly in the  remainder of the slice could support this argument. However, the evidence does not conclusively support this interpretation either because there is no discontinuity in the variation of the azimuth like can be observed in the magnetic loop shown in the second row of Fig.~\ref{fig:ROIs}. We note that without disambiguation of the azimuth we cannot be certain that the gradual, coherent variation we measure is correct. If this alternative explanation was true we argue that we should not observe linear polarisation signals at the middle, apparent PIL between the magnetic loops. In other words, we should not observe linear polarisation signals at all three apparent PILs. We note that linear polarisation signals are pervasive across the entire slice, and thus we are able to observe two crests in the magnetic field lines. However, we also note that we did discover a second serpentine structure with three PILs in our analysis, and the middle PILs did not have linear polarisation signals. 

Ultimately, only repeated observations with a high-cadence time series would allow us to conclusively determine the correct scenario. Finally we point out that the temporal resolution may play a role due to the scanning speed of the observations. The step cadence of these scans is about $7.4$ seconds. A horizontal flow would have to be on the order of $4$~km sec$^{-1}$ to be similar to the stepping cadence. However, it would be premature to say this is a factor which makes either scenario less or more plausible; for instance,  the gradual variation in the azimuth that we measured could be explained by supposing that we observed the temporal evolution of a serpentine structure.

A physical scenario that demands an increased number of free parameters commonly occurs at the PIL of magnetic elements, and detecting linear polarisation at these locations has been a major challenge \citep{kubo2010,kubo2014}. We examined two magnetic loops which had very different properties at the PIL in terms of circular polarisation, but which both had strong linear polarisation signals. In one case, we uncovered a Stokes vector which had a complex Stokes $V$ profile. Three-lobed Stokes $V$ profiles were also observed in the quiet Sun by \cite{Viticchi2011} with Hinode/SP and \cite{marian2016,kiess2018,Campbell2021a} with GREGOR. Initially we attempted a 2C inversion, but with gradients in $\gamma$ as well as in $B$ and $v_{\mathrm{LOS}}$ (see Fig.~\ref{fig:profileB1}). We encountered a problem of degeneracy as we found this profile could also be explained with a more extreme gradient in $\gamma$ but without two magnetic model atmospheres (see Fig.~\ref{fig:profileB1}). The second interpretation is consistent with \cite{khomenko2005}, who showed using magnetoconvection simulations that gradients in $\gamma$ and $v_{\mathrm{LOS}}$ can generate irregular Stokes $V$ profiles. In this case it is not as simple as saying the approach with the lower number of free parameters is the correct one - the fact that this profile was located along the PIL means it is plausible that the 3-lobed Stokes $V$ profile is the physical manifestation of a smearing of signals of opposite polarities, or even mixing within the spatiotemporal resolution element of the observations that remains unresolved. We note that in the 1C case $\gamma$ changes polarity along the LOS in the optical depth range in which these lines are sensitive to perturbations in $\gamma$, and there are other examples in this study where this occurs (see Fig.~\ref{fig:profileA} and Fig.~\ref{fig:profileC1}). Previously \cite{rez2007} had shown this was possible when the Fe I $6301.5/6302.5$~$\mathrm{\AA}$ doublet showed two-lobed Stokes $V$ profiles with different polarities in a single spectrum, but in our observations we find the Stokes $V$ profiles of both lines are compatible. 

In another case, we observe a long, narrow magnetic loop at the granule-IGL boundary which has an absence of Stokes $V$ at the PIL. It is most plausible that in this case the magnetic field at the PIL is completely transverse, however, like in the former example, opposite polarity Stokes $V$ signals may have mixed such that their signals have completely cancelled each other. However, the absence of a Stokes $V$ signal cannot be presented as evidence for the latter scenario. We also find degenerate solutions to this Stokes vector, based on a single magnetic model and two magnetic models (see Fig.~\ref{fig:profileB2}), showing that degeneracy is an issue even in the absence of a Stokes $V$ profile. This degeneracy issue is a serious problem which calls into question how accurately we can infer physical parameters from inversions, but we argue this issue could be significantly improved by observing more photospheric spectral lines. This is distinct from the degeneracy problem described by \cite{marian2006} because in all the inversions presented in this study we forced the temperature of each component to be the same. Ultimately, if the solutions requiring two magnetic models are correct it suggests that, even at the spatiotemporal resolution of DKIST/ViSP, magnetic flux is yet hidden to spectropolarimetric observations that use the Zeeman effect in magnetic loops (or bi-poles). 

Significant effort was made to try and balance the need for an increased number of free parameters to fit complex Stokes profiles with the risk of producing unphysical or unrealistic solutions from the inversions. In particular, velocity gradients in the LOS often manifest in Stokes $Q$, $U$, and $V$ as an amplitude asymmetry that cannot be modelled with a $v_{\mathrm{LOS}}$ stratification that is constant in optical depth. In the final case study we examined a magnetic feature spanning a granule which had linear polarisation signals throughout most of its structure. The magnetic nature of granules is important because it has implications for constraining magnetohydrodynamic simulations, particularly in terms of circulation models, emergence of magnetic flux, and the formation of IGLs \citep{rempel2018}. It is notable that this granule has pervasive linear polarisation, in addition to circular polarisation, because magnetic flux emergence in granules was observed in early Hinode/SP observations to be longitudinal in nature \citep{granules2008}, but the emergence of horizontal magnetic flux has been observed in a granule \citep{gom2010}. In the granule, we found symmetric Stokes profiles (see Fig.~\ref{fig:profileC2}), but at the granule-IGL boundary we sampled a Stokes vector from a pixel which had significant asymmetries in the polarisation parameters (see Fig.~\ref{fig:profileC1}). We inverted this Stokes vector with an approach which sought to minimise the number of free parameters as much as possible, but were only adequately able to reproduce it using gradients in $B$, $\gamma$, and $v_{\mathrm{LOS}}$. Single-lobed Stokes $V$ profiles are common in quiet Sun Hinode observations, accounting for up to $34\%$ of profiles \citep{Viticchi2011}. They cover about $2\%$ of the surface, are associated with inclined magnetic fields and flux emergence and submergence processes, and are also reproduced by simulations \citep{singlelobed2012}. There is an expectation that the existence of canopy fields mean that the magnetic field becomes statistically more horizontal with height \citep{stenflo2013,danilovic2016}. However, in our analysis we found the opposite in individual cases (see Fig.~\ref{fig:profileA}~and~\ref{fig:profileC1}). Although it is well understood that depth-averaged parameters can be well retrieved from inversions of photospheric diagnostics without the inclusion of gradients in the magnetic or kinetic parameters \citep[see, for instance,][]{Campbell2021b,Carlos2023}, we offer a word of caution that at DKIST resolutions it is clear that any future statistical analysis of this data will be inadequate unless the need for gradients in $v_{\mathrm{LOS}}$ (and other parameters) are accounted for in the inversions due to their prevalence in these small-scale magnetic structures.

We sampled relatively symmetric circular polarisation profiles at the granular boundaries of the magnetic granule - with one side of the granule harbouring smaller amplitude profiles, indicative of a weak magnetic field, and the other with amplitudes twice their size and much larger lobe separations - and applied the weak and strong field approximations, to retrieve manual estimates of the $\alpha_m B$ and $B$. This allowed us to sanity check our measurements of vastly different magnetic field strengths spanning two orders of magnitude. However, we note that the difference in $\alpha_m B$ is not as large as the difference in $B$, indicating that the determination of $\alpha_m$ plays a role. However, we expect that inversions of these lines are able to constrain both the $\alpha_m$ and $B$ \citep{ramos2009} and expect to be able to make distinctions between weak and strong fields in these lines when $\alpha_m$ is a free parameter \citep{iniesta2010}. We emphasise that such symmetrical Stokes $V$ profiles are in the minority in this structure, especially in the IGLs, and so we do not recommend attempting to apply these approximations more broadly to these data except in limited specific cases. The extremely large field strengths we recorded are at the high end of what is expected to result from magnetic field amplification in bright points based on high resolution spectropolarimetric observations \citep[see, for instance,][]{beck2007,utz2013,keys2019}.

The analysis presented in this study is conducted under the assumption of local thermodynamic equilibrium (LTE). \cite{nlte} have shown that for the Fe I $6300$~$\mathrm{\AA}$ doublet there can be departures from LTE due to resonance line scattering and over-ionisation of Fe atoms by ultraviolet photons. The former process makes the line stronger, while the latter makes the line weaker, which means for the Fe I $6300$~$\mathrm{\AA}$ doublet these processes compensate each other to an extent.  Nevertheless, departures from LTE can introduce errors or uncertainties in the derived atmospheric parameters. 

\section{Conclusions}\label{sect:conclusions}
We have analysed spectropolarimetric observations of the quiet Sun with the first $4-$m class optical telescope in service to solar physics. This allowed us to reveal, for the first time, the serpentine nature of the sub-granular, small-scale magnetic field, as we demonstrated that the polarity of the magnetic field changed at least three times in only $400$~km. The three polarity reversals are evident from an inspection of the Stokes $V$ profiles and confirmed by the inversions. From the inversions we were also able to deduce that the azimuth changed in a significant but coherent way. While beyond the scope of this study, it is clear that a statistical analysis based on an non-LTE inversion, not just of the Ca II $8542$~$\mathrm{\AA}$ line but also the Fe I doublet, is the next logical step and will be the subject of future work. We were unable to trace the temporal evolution of these structures. Observations of these structures at high-cadence ($30-90$ seconds) should be a priority for DKIST in the future with the Diffraction Limited Near Infrared Spectropolarimeter (DL-NIRSP) instrument \citep{dlnirsp}.

We found that, particularly at PILs of magnetic loops, degenerate solutions in inversions remains a significant issue even at the spatial resolution provided by a 4-m telescope, and this is true even in the absence of a Stokes $V$ signal. Although previous work involving simulations and synthetic observations have shown that atmospheric parameters can be well recovered on average (see, for instance, \cite{Campbell2021b, Carlos2023}), in our case studies we found that to truly characterise the plasma on small scales we had to permit a larger number of free parameters to produce the required asymmetries in the Stokes profiles and abnormal Stokes $V$ profiles. We are unable to distinguish between gradients in atmospheric parameters along the LOS and smearing of signals in the plane of the solar surface due to the spatial point spread function of the telescope. We concur with the advice provided by \cite{cabrera2005,quintero2021} and suggest that observations of the Fe I $6302.5$~$\mathrm{\AA}$ line with the deep photospheric Fe I $15648.5$~$\mathrm{\AA}$ line should be prioritised, as observations of diagnostics which sample a range of heights in the photosphere could provide a method for distinguishing between these two otherwise plausible scenarios when simultaneously inverted. These observations would additionally benefit from the demonstrated ability of the Fe I $15648.5$~$\mathrm{\AA}$ line to observe higher linear polarisation fractions \citep{marian2008_vis_NIR,lagg2016,campbell2023}.

\begin{acknowledgments}
We thank the anonymous referee for their suggestions which improved this manuscript. RJC, DBJ and MM acknowledge support from the Science and Technology Facilities Council (STFC) under grant No. ST$/$P000304$/$1,  ST$/$T001437$/$1, ST$/$T00021X$/$1  and ST$/$X000923$/$1. DBJ acknowledges support from the UK Space Agency for a National Space Technology Programme (NSTP) Technology for Space Science award (SSc~009). DBJ and MM acknowlege support from The Leverhulme Trust grant RPG-2019-371. 
This research has received financial support from the European Union's Horizon 2020 research and innovation program under grant agreement No. 824135 (SOLARNET). The research reported herein is based in part on data collected with the Daniel K. Inouye
Solar Telescope (DKIST), a facility of the National Solar Observatory (NSO). NSO is managed by the Association
of Universities for Research in Astronomy, Inc., and is funded by the National Science Foundation. 
Any opinions, findings and conclusions or recommendations expressed in this publication are those of the
authors and do not necessarily reflect the views of the National Science Foundation or the Association of
Universities for Research in Astronomy, Inc. DKIST is located on land of spiritual and cultural significance
to Native Hawaiian people. The use of this important site to further scientific knowledge is done so with
appreciation and respect. This material is based upon work supported by the National Center for Atmospheric Research, which is a major facility sponsored by the National Science Foundation under Cooperative Agreement No. 1852977.
\end{acknowledgments}

The observational data used during this research is openly available from the \hyperlink{https://dkist.data.nso.edu}{DKIST Data Centre Archive} under proposal identifier pid$\_1\_36$.

\textbf{\facilities{DKIST \citep{DKIST2020}.}}

\textbf{\software{\hyperlink{https://github.com/r-j-campbell/SIRExplorer}{SIR explorer} \citep{SIRE}, Astropy \citep{astropy3}, Matplotlib \citep{mpl}, Numpy \citep{numpy}.}}


\bibliography{sample631}{}
\bibliographystyle{aasjournal}

\end{document}